\documentclass[12pt,aps]{revtex4}
\usepackage{amsmath,amssymb,graphicx,mathrsfs}
\newcommand{\be}{\begin{equation}}
\newcommand{\ee}{\end{equation}}
\newcommand{\bea}{\begin{eqnarray}}
\newcommand{\eea}{\end{eqnarray}}

\def\({\left(} \def\){\right)}

\def\revise#1       {\raisebox{-0em}{\rule{3pt}{1em}}
                     \marginpar{\raisebox{.5em}{\vrule width3pt\
                     \vrule width0pt height 0pt depth0.5em
                     \hbox to 0cm{\hspace{0cm}{%
                     \parbox[t]{4em}{\raggedright\footnotesize{#1}}}\hss}}}}

\begin{document}
\title{Wald's entropy is equal to a quarter of the horizon area in units of the effective gravitational coupling}
\author{Ram Brustein}
\affiliation{Department of Physics, Ben-Gurion University,
    Beer-Sheva 84105, Israel\\ E-mail: ramyb@bgu.ac.il }
 \author{   Dan Gorbonos } \affiliation{ Department of Physics, University of Alberta,
Edmonton, Alberta, Canada T6G 2G7 \\
    E-mail: gorbonos@phys.ualberta.ca}
  \author{  Merav Hadad} \affiliation {Department of Natural Sciences, The Open University of Israel,
P.O.B. 808, Raanana 43107, Israel \\  and
School of Engineering, The Ruppin Academic Center,
Emeq Hefer 40250, Israel
 \\  and
Department of Physics, Ben-Gurion University,
    Beer-Sheva 84105, Israel\\ E-mail: meravha@openu.ac.il}


\begin{abstract}
The Bekenstein-Hawking entropy of black holes in Einstein's theory of gravity is equal
to a quarter of the horizon area in units of Newton's constant. Wald
has proposed that in general theories of gravity  the entropy of
stationary black holes with bifurcate Killing horizons is a Noether
charge which is in general different from the Bekenstein-Hawking
entropy. We show that the Noether charge entropy is equal to a
quarter of  the horizon area in units of the effective gravitational
coupling on the horizon defined by the coefficient of the kinetic
term of a specific metric perturbation polarization on the horizon. We present
several explicit examples of static spherically symmetric black
holes.
\end{abstract}


\maketitle
\preprint{}


\section{Introduction}
\label{intro}

The Bekenstein-Hawking entropy of black holes (BH's) in Einstein's
theory of gravity is equal to a quarter of the horizon area in units
of Newton's constant \cite{bekenstein,hawking}. Wald
\cite{wald1,wald2} has studied BH's in generalized theories of
gravity and proposed that the correct dynamical entropy of
stationary BH's solutions with bifurcate Killing horizons is a
Noether charge entropy.

The Noether charge entropy is in general different from the Bekenstein-Hawking entropy. First, the Noether charge entropy is local: it can be defined in terms of quantities on the horizon. Further, the Noether charge entropy was found to be invariant under field redefinitions that do not change the structure of space-time at infinity and on the horizon \cite{jacobson}.  In Einstein's gravity there is only one dimensional parameter $G_N$ and from it (and $\hbar$ and the speed of light $c$) it is possible to construct a single parameter with units of length, the Planck length $l_P^2=\hbar G_N/c^3$. In more general theories additional parameters can appear and hence several length scales can replace $l_P$.

The validity of Wald's proposal has been checked in many examples in a string theory context where the direct counting of microstates can be compared explicitly to the Noether charge entropy \cite{Lopes Cardoso:1998wt}. To the best of our knowledge all the explicit comparisons were done for static solutions or those that are equivalent to static solutions. Unfortunately, stationary solutions for which the corrections to the Einstein-Hilbert action are significant are not known, so an explicit comparisons could not be done for non-static solutions. An early review of the subject can be found in \cite{Mohaupt:2000mj} and a recent and much more extensive review can be found in \cite{sen2007}.

Our goal in this paper was to clarify the relationship between the Noether Charge
entropy and the Bekenstein-Hawking entropy. Our motivation was to
resolve the apparent tension between the entanglement interpretation
of BH entropy and the Noether charge entropy \cite{waldliving}, to
understand its geometrical dependence and to explain some
calculation of the entropy in string theory \cite{giveon1,giveon2}
in which the entropy of charged BH's with higher derivative corrections was found to depend on the charges only through the horizon area. Previously, it was observed in \cite{giveon3} that the entropy of two dimensional charged BH's is proportional to the area of the horizon for any value of the charges and the mass.

We have discovered that the Noether charge entropy is equal to a
quarter of  the horizon area in units of the effective gravitational
coupling on the horizon rather than in units of $G_N$. The effective gravitational
coupling on the horizon is defined by the coefficient of the kinetic
term of a specific metric perturbation polarization on the horizon. In Einstein's gravity both definitions coincide, however in general they do not.  We discuss
several explicit examples of static spherically symmetric black
holes.

The rest of the paper is organized as follows. In section~\ref{wald} we review the Noether charge entropy, in section~\ref{kinetic} we recall the definition of the effective gravitational coupling and show that it is equal to the functional derivative of the Lagrangian density with respect to  the Riemann tensor. In section~\ref{noether} we discuss our main result and show that the Noether charge entropy is equal to a
quarter of  the horizon area in units of the effective gravitational
coupling on the horizon. In section~\ref{polariz} we identify the metric perturbation polarizations chosen in the Wald formula as those associated with the perturbations of the area density on the bifurcation surface. In section~\ref{examples} we present several explicit examples of entropy and gravitational coupling and verify our results. Section~\ref{discussion} contains a discussion of our result and its significance and an outlook.

\section{The Noether charge entropy}
\label{wald}

A general theory of gravity whose action depends on the metric $g_{\mu\nu}$, the curvature (through the Riemann tensor) and matter fields $\phi$ and their covariant derivatives
\begin{equation}
\label{genlag}
I=\int d^D x \sqrt{-g}\ \mathscr{L}\left(R_{\rho\mu\lambda\nu},g_{\mu\nu},\nabla_\sigma R_{\rho\mu\lambda\nu},\phi,\nabla\phi,\ldots\right),
\end{equation}
may have stationary BH solutions with bifurcate killing horizons. According to Wald \cite{wald1,wald2}, the Noether charge entropy for such BH's is
\begin{equation}
\label{waldentropygen}
S_{W}=-2 \pi \oint\limits_{\Sigma} \left(\frac{\delta\mathscr{L}}{\delta R_{abcd}}\right)^{\!\!(0)} \hat\epsilon_{ab}\hat\epsilon_{cd}\bar{\epsilon}.
\end{equation}
The Noether charge entropy was first expressed in this form in \cite{jacobson}.
If derivatives of the Riemann tensor appear in $\mathscr{L}$ then one is to perform an integration by parts first and then take the derivative. The procedure is similar to finding the Euler-Lagrange equations in a theory with higher derivatives of the canonical variables.

The integral in eq.~(\ref{waldentropygen}) is on the $D-2$ dimensional space-like bifurcation surface $\Sigma$. The hatted variable $\hat\epsilon_{ab}$ is the binormal vector to
the bifurcation surface. It is antisymmetric under the exchange
$a\leftrightarrow b$ and  normalized as
$\hat\epsilon^{ab}\hat\epsilon_{ab}=-2$. This normalization sets the
computation of the entropy in units such that the BH temperature is
$\frac{1}{2\,\pi}$ (see~\cite{wald1} for details). The variable
$\bar{\epsilon}$ is the induced volume form on the
bifurcation surface.
The superscript $(0)$ indicates that the partial derivative
$\left(\frac{\delta\mathscr{L}}{\delta R_{abcd}}\right)^{\!\!(0)}$
is evaluated on the solution of the equations of motion. The variation of the Lagrangian with respect to
$R_{abcd}$ is performed  as if $R_{abcd}$ and the metric $g_{\mu\nu}$ are independent and
it includes contributions from the covariant derivatives acting on
matter fields. The covariant derivatives have to be expressed as symmetric and
antisymmetric combinations and then they have to be expressed
in terms of the Riemann tensor (See section 2 of \cite{wald2} for a detailed explanation).

Since our examples will be of static BH's we write all the expressions explicitly for this case.  For static spherically symmetric BH solutions in $D=d+1$ space-time dimensions that posses a bifurcate Killing horizon the metric can be brought to a canonical form,
\begin{equation}
\label{BHmetric}
ds^2=- f(r) dt^2 + \frac{1}{f(r)} dr^2 + q(r) d\Omega_{d-1}^2.
\end{equation}
The function $f(r)$ vanishes at the event horizon $r=r_H$, the bifurcation surface is at $r=r_H$, $t=const.$ and $d\Omega_{d-1}^2$ is the spherical volume element.

For these BH's the relevant Killing vector is $\partial_t$ and  $\hat\epsilon_{tr}=1$.
The $\hat\epsilon$'s vanish for  $a,b\ne t,r$. The explicit
expression for the Noether charge entropy is
\begin{eqnarray}
\label{waldexplicit}
S_{W}&=&-2 \pi \oint\limits_{r=r_H, t=const.} \left(\frac{ \delta \mathscr{L}}{ \delta R_{abcd}}\right)^{(0)}\hat\epsilon_{ab}\hat{\epsilon}_{cd} \left[q(r)\right]^{\frac{d-1}{2}} d\Omega_{d-1}^2 \nonumber \\
&=&-8 \pi \oint\limits_{r=r_H, t=const.} \left(\frac{ \delta \mathscr{L}}{ \delta R_{rtrt}}\right)^{(0)}\left[q(r)\right]^{\frac{d-1}{2}} d\Omega_{d-1}^2.
\end{eqnarray}
The factor of 4 come from the antisymmetry properties of the Riemann tensor and the binormal vectors. The superscript $(0)$ emphasizes that the functional derivative is evaluated on the solution.

A few examples will be useful. First, let us see how the Noether
charge entropy reproduces the Bekenstein-Hawking area entropy for
the Einstein-Hilbert (EH) action
$\mathscr{L}_{EH}=\frac{1}{16\pi G_N} R$,
\begin{eqnarray}
S_{W}&=&-8 \pi \oint\limits_{r=r_H, t=const.} \frac{1}{16\pi G_N} \left[\left(\frac{ \delta R}{ \delta R_{rtrt}}\right)^{(0)} \right] \left[q(r)\right]^{\frac{d-1}{2}} d\Omega_{d-1}^2  \nonumber \\
&=& -\frac{1}{4 G_N} \oint\limits_{r=r_H, t=const.}   \left(\overline{g}^{tt}\overline{g}^{rr}- \overline{g}^{tr}\overline{g}^{tr}\right) \left[q(r)\right]^{\frac{d-1}{2}} d\Omega_{d-1}^2  \nonumber \\
&=& \frac{A_H}{4 G_N}.
\end{eqnarray}
We have denoted the background metric solution by
$\overline{g}_{\mu\nu}$ and used the fact that it is of the form
(\ref{BHmetric}) for which $\overline{g}^{tt}\overline{g}^{rr}=-1$
and $\overline{g}^{tr}=0$. The area of the horizon $A_H$ is given by
$A_H=\oint\limits_{r=r_H, t=const.}\left[q(r)\right]^{\frac{d-1}{2}}
d\Omega_{d-1}^2$.

As a second example let us consider dilaton-gravity $\mathscr{L}=\frac{e^{-2\phi}}{16\pi
G_N}  R$ and assume that the solution is spherically symmetric
$\phi=\phi(r)$. Then
\begin{eqnarray}
S_{W}&=&-8 \pi \oint\limits_{r=r_H, t=const.} \frac{ e^{-2\phi(r)} }{16\pi G_N} \left[\left(\frac{ \delta R}{ \delta R_{rtrt}}\right)^{(0)} \right] \left[q(r)\right]^{\frac{d-1}{2}} d\Omega_{d-1}^2  \nonumber \\
&=& -\frac{1}{4 G_N} \oint\limits_{r=r_H, t=const.}  e^{-2\phi(r)} \left(\overline{g}^{tt}\overline{g}^{rr}- \overline{g}^{tr}\overline{g}^{tr}\right) \left[q(r)\right]^{\frac{d-1}{2}} d\Omega_{d-1}^2  \nonumber \\
&=& \frac{A_H}{4 e^{2\phi(r_H)}G_N}.\label{entropy-dilaton}
\end{eqnarray}
Clearly in the case of dilaton-gravity the parameter $G_N$ does not
determine by itself the strength of the gravitational coupling or of
the gravitational force, rather they are determined the combination
$e^{2\phi(r)}G_N$ which can depend on $r$. Now consider evaluating
the Noether charge entropy in the Einstein frame. The Einstein-frame
metric is $\hat{g}_{\mu\nu}=e^{-\frac{4}{d-1} \phi} g_{\mu\nu}$, and
in the Einstein frame the Lagrangian density is
$\mathscr{L}=\frac{1}{16\pi G_N}  \widehat{R}+\cdots$. Then
\begin{eqnarray}
S_{W}&=&-8 \pi \oint\limits_{r=r_H, t=const.} \frac{1}{16\pi G_N} \left[\left(\frac{ \delta \widehat{R}}{ \delta \widehat{R}_{rtrt}}\right)^{(0)} \right] \left[ \widehat{q}(r)\right]^{\frac{d-1}{2}} d\Omega_{d-1}^2  \nonumber \\
&=& -\frac{1}{4 G_N} \oint\limits_{r=r_H, t=const.}   \left(\overline{\widehat{g}}{}^{tt}\overline{\widehat{g}}{}^{rr}- \overline{\widehat{g}}{}^{tr}\overline{\widehat{g}}{}^{tr}\right) \left[ \widehat{q}(r)\right]^{\frac{d-1}{2}} d\Omega_{d-1}^2  \nonumber \\
&=& \frac{\widehat{A}_H}{4 G_N}.
\end{eqnarray}
Since $\widehat{q}(r)=e^{-\frac{4}{d-1}\phi(r)}q(r)$, then, as anticipated, the entropies evaluated in both frames are equal $$ \frac{\widehat{A}_H}{4 G_N} = \frac{A_H}{4 e^{2\phi(r_H)}G_N}.$$

\section{The effective gravitational coupling and the metric perturbations kinetic terms}
\label{kinetic}

Here we discuss the gravitational coupling and the
kinetic terms of metric perturbations for a general background and not necessarily
for a BH background. We first recall the definition of the
gravitational coupling in Einstein's theory. One expands the metric
about a fixed background solution $\overline{g}_{\mu\nu}$,
\begin{equation}
 g_{\mu\nu}=\overline{g}_{\mu\nu}+h_{\mu\nu}.
\end{equation}
The inverse metric is:
\begin{equation}
 g^{\mu\nu}=\overline{g}^{\mu\nu}-h^{\mu\nu}
\end{equation}
and the indices of $h_{\mu\nu}$ are raised and lowered with
the background metric. The action can be expanded in powers of $h_{\mu\nu}$. The equations of motion imply that the linear term in this expansion vanishes.

The expansion of the Einstein-Hilbert Lagrangian $\mathscr{L}_{EH}$ in $h_{\mu\nu}$ to second order~\cite{dono} is
\begin{equation}
\frac{1}{16\,\pi\,G}\,\sqrt{-g}\,R=\frac{1}{16\,\pi\,G}\, \sqrt{-\overline{g}}\left(\overline{R}+\mathscr{L}_{EH}^{(2)}\right),
\end{equation}
with
\begin{eqnarray}
\label{ehexpansion}
\mathscr{L}_{EH}^{(2)}&=& \frac{1}{4}\,\overline{\nabla}_{\alpha}\,h_{\mu\nu}\, \overline{\nabla}^{\alpha}\,h^{\mu\nu} -\frac{1}{4}\,\overline{\nabla}_{\alpha}h\, \overline{\nabla}^{\alpha}h+ \frac{1}{2} \overline{\nabla}_{\alpha} h\overline{\nabla}_{\beta}\,h^{\alpha\beta} \nonumber\\
&& -\frac{1}{2}\overline{\nabla}_{\alpha}\,h_{\mu\beta}\,\overline{\nabla}^{\beta}\,h^{\mu\alpha} +\overline{R}\,\left(\frac{1}{4}\,h^{2} -\frac{1}{4}\,h_{\mu\nu}\,h^{\mu\nu}\right)\nonumber\\
&&+\overline{R}^{\mu\nu}\,\left(\,h_{\ \mu}^{\alpha}\,h_{\nu\alpha}-\frac{1}{2} h\,h_{\mu\nu}\right),
\end{eqnarray}
where $h=h_{\ \lambda}^{\lambda}$.
For a background metric that solves the vacuum Einstein equations the last two terms vanish.

We now wish to look at the kinetic terms (terms with two derivatives) of the metric perturbations in the expansion
\begin{eqnarray}
\frac{1}{4}\left(\,\overline{\nabla}_{\alpha}\,h_{\mu\nu}\, \overline{\nabla}^{\alpha}\,h^{\mu\nu}- 2\overline{\nabla}_{\alpha}\,h_{\mu\beta}\,\overline{\nabla}^{\beta}\,h^{\mu\alpha} -\overline{\nabla}_{\alpha}h\, \overline{\nabla}^{\alpha}h+ 2 \overline{\nabla}_{\alpha} h\overline{\nabla}_{\beta}\,h^{\alpha\beta}  \right). \nonumber
\end{eqnarray}
We can determine the gravitational coupling $\kappa^2=32 \pi G_N$ from the kinetic terms,
\begin{eqnarray}
\label{kappadef}
&&\frac{1}{64\,\pi\,G}\,\left(\,\overline{\nabla}_{\alpha}\,h_{\mu\nu}\, \overline{\nabla}^{\alpha}\,h^{\mu\nu}- 2\overline{\nabla}_{\alpha}\,h_{\mu\beta}\,\overline{\nabla}^{\beta}\,h^{\mu\alpha} -\overline{\nabla}_{\alpha}h\, \overline{\nabla}^{\alpha}h+ 2 \overline{\nabla}_{\alpha} h\overline{\nabla}_{\beta}\,h^{\alpha\beta}  \right)  \nonumber \\ =&& \frac{1}{2}\frac{1}{\kappa^2}\ \ \ \left(\,\overline{\nabla}_{\alpha}\,h_{\mu\nu}\, \overline{\nabla}^{\alpha}\,h^{\mu\nu}- 2\overline{\nabla}_{\alpha}\,h_{\mu\beta}\,\overline{\nabla}^{\beta}\,h^{\mu\alpha} -\overline{\nabla}_{\alpha}h\, \overline{\nabla}^{\alpha}h+ 2 \overline{\nabla}_{\alpha} h\overline{\nabla}_{\beta}\,h^{\alpha\beta}  \right).
\end{eqnarray}

The coefficients matrix of the kinetic terms is not diagonal in the metric perturbations so to identify correctly the gravitational couplings it needs to be diagonalized. The eigenvectors $H^i_{\mu\nu}$ are given by linear combinations of the original metric perturbations $h_{\mu\nu}$.  To verify that $\kappa$ in eq.(\ref{kappadef}) is truly the gravitational coupling one expands $g_{\mu\nu}=\overline{g}_{\mu\nu}+\kappa H_{\mu\nu}$. Then the kinetic term for the metric perturbations $H_{\mu\nu}$ becomes canonical and each factor of $H_{\mu\nu}$ in the interaction terms is accompanied by a factor of $\kappa$. The most general coefficients matrix of the kinetic terms is a six index object. However, as can be seen from eq.(\ref{kappadef}), due to symmetries it is actually a four index object. When the background spacetime is symmetric under rotations, the different helicities of the metric perturbations can be further separated into independent tensors, vectors and scalars and the coefficient matrix of the kinetic terms can be diagonalized.  Some of the metric perturbations are gauge degrees of freedom that can be removed by an appropriate choice of coordinates. Obviously, some degrees of freedom of the metric perturbations are physical either on their own sake or by mixing with matter degrees of freedom.

The gravitational coupling in a general theory of gravity can be determined in a similar way. We take a general action
\begin{equation}
I=\int d^D x \sqrt{-g}\ \mathscr{L}\left(R_{\rho\mu\lambda\nu},g_{\mu\nu},\nabla_\sigma R_{\rho\mu\lambda\nu},\phi,\nabla\phi,\ldots\right)
\end{equation}
and expand the metric $ g_{\mu\nu}=\overline{g}_{\mu\nu}+h_{\mu\nu}.$ The action can be expanded $I=I^{(0)}+\delta I^{(1)}+ \delta I^{(2)}+\cdots$. We are interested in contributions to terms in the effective action of the metric perturbations that are quadratic in the perturbations and quadratic in derivatives. We call such terms ``kinetic terms". The most general coefficients matrix of the kinetic terms is a six index object. However, as we will show, due to symmetries it is actually a four index object also in the general case.

As in the Einstein case, the coefficients matrix of the kinetic terms is not diagonal in the metric perturbations. Obviously, since the general theory contains additional couplings, the coefficients matrix can have different eigenvalues for different $H_{\mu\nu}$'s leading to different gravitational couplings $\kappa^i$ for each $H_{\mu\nu}^i$.
In the basis in which the kinetic terms are diagonal one then expands $g_{\mu\nu}=\overline{g}_{\mu\nu}+\kappa^i H^i_{\mu\nu}$. In this general case we may also verify that $\kappa^i$ may be called  ``gravitational couplings". The kinetic term for any of the polarizations in this basis  becomes canonical and  each factor of $H^i_{\mu\nu}$ in the interaction terms is accompanied by a factor of $\kappa^i$. A general action may contain additional dimensionless or dimensionful  couplings that determine the strength of a particular type of interaction. However,  all the interactions of a specific polarization can be classified according to their overall power of the appropriate $\kappa^i$. All interactions will vanish in the limit $\kappa^i \to 0$ when the other couplings are held fixed.

From general covariance it follows that derivatives of $h_{\mu\nu}$ can only appear through
the Christoffel symbols either in combinations involving the Riemann
tensor and its derivatives, or through covariant derivatives of
matter fields. The covariant derivatives of matter fields
can be expressed as symmetric and antisymmetric combinations.
Both symmetric and antisymmetric contributions can be expressed
in terms of the Riemann tensor and its derivatives as explained in \cite{wald2}.
The argument is most transparent in Riemann Normal coordinates in which the Christoffel symbols vanish locally and derivatives of the Christoffel symbols can be expressed in terms of the Riemann tensor and its derivatives.
Once the derivatives of $h_{\mu\nu}$ are expressed in terms of covariant background tensors then the action built from them is invariant under general background coordinate transformations so the conclusion that the derivatives of $h_{\mu\nu}$ appear only through the the Riemann tensor and its derivatives holds for any coordinate system.  We
therefore know that contributions to the metric perturbation kinetic terms must
appear only through factors of the Riemann tensor (or its derivatives) in
the action.

Our goal is to find the coefficients matrix of the kinetic terms. We have just argued that the kinetic terms appear only through factors of the Riemann tensor (and its derivatives). It follows that we need to focus on the specific contribution to  the variation $\delta I$
\begin{equation}
\delta I \sim\int d^D x \sqrt{-g} \frac{
 \delta \mathscr{L}}{ \delta R_{\rho\mu\lambda\nu}}\delta
R_{\rho\mu\lambda\nu}
\end{equation}
and look in the expansion for terms that contain two factors of the metric perturbation
and two background covariant derivatives.

The variation of the Riemann tensor can be expressed as
\begin{equation}
\label{Rvariation2}
\delta R_{\rho\mu\lambda\nu}={\nabla}_\lambda\delta\Gamma_{\nu\mu\rho}
-{\nabla}_\nu\delta\Gamma_{\lambda\mu\rho}.
\end{equation}
Consequently, the relevant variation is
\begin{equation}
\delta\mathscr{L}=\frac{\delta \mathscr{L}}{ \delta
R_{\rho\mu\lambda\nu}}\left({\nabla}_\lambda\delta\Gamma_{\nu\mu\rho}
-{\nabla}_\nu\delta\Gamma_{\lambda\mu\rho}\right).
\end{equation}
In principle we could imagine expanding all factors in the metric perturbation to obtain the contribution to the kinetic terms. However, as we will now show, we need to expand only the second factor $\left({\nabla}_\lambda\delta\Gamma_{\nu\mu\rho}
-{\nabla}_\nu\delta\Gamma_{\lambda\mu\rho}\right)$.

The expansion of ${\nabla}_\lambda\delta\Gamma_{\nu\mu\rho}
-{\nabla}_\nu\delta\Gamma_{\lambda\mu\rho}$ in $h_{\mu\nu}$ contains
at least two derivatives and at least one factor of the perturbation:
\begin{equation}
{\nabla}_\lambda\delta\Gamma_{\nu\mu\rho}
-{\nabla}_\nu\delta\Gamma_{\lambda\mu\rho}=\overline{\nabla}_\lambda\Upsilon_{\nu\mu\rho}
-\overline{\nabla}_\nu\Upsilon_{\lambda\mu\rho}+\Upsilon_{\lambda\delta\rho}\Upsilon_{\nu\mu}^{\
\ \delta} -\Upsilon_{\delta\nu\rho}\Upsilon_{\mu\lambda}^{\ \
\delta}+{\cal O}\(h^{2}\).
\label{expansion}
\end{equation}
Here we have denoted for convenience
\begin{equation}
\delta\Gamma^{(1)}_{\alpha\beta\gamma}\equiv\Upsilon_{\alpha\beta\gamma}= \frac{1}{2}\left(\overline{\nabla}_{\alpha}h_{\beta\gamma} + \overline{\nabla}_{\beta}h_{\alpha\gamma} -\overline{\nabla}_{\gamma}h_{\alpha\beta}\right).
\end{equation}
Hence terms in the expansion of $\sqrt{-g}\frac{\delta \mathscr{L}}{ \delta R_{\rho\mu\lambda\nu}}$ can contribute to kinetic terms only if they contain exactly one factor of $h_{\mu\nu}$ without derivatives. Such terms can come only from the linear term in the expansion of $g^{\mu\nu}$,
\begin{equation}
g^{\mu\nu}=\overline{g}^{\mu\nu}-h^{\mu\nu}+{\cal O}\(h^{2}\).
\end{equation}

However, if we had a term which contained a $g^{\mu\nu}$ in $\sqrt{-g}\frac{\delta
\mathscr{L}}{ \delta R_{\rho\mu\lambda\nu}}$ it could have been
canceled by integration by parts because $\nabla_\lambda g^{\mu\nu}=0$:
\begin{equation}
\int d^D x \sqrt{-g}\delta\mathscr{L}\sim-2 \int d^D x \,{\nabla}_\lambda\left(\sqrt{-g}\frac{\delta \mathscr{L}}{
\delta R_{\rho\mu\lambda\nu}}\right)\,\delta\Gamma_{\nu\mu\rho}.
\end{equation}

We conclude that we need to look only at the terms which are second order
in $h_{\mu\nu}$ in the expansion of $\delta R_{\rho\mu\lambda\nu}$
and zeroth order in $\sqrt{-g}\frac{\delta \mathscr{L}}{ \delta
R_{\rho\mu\lambda\nu}}$. In other words, kinetic terms can appear
only through $\delta\mathscr{L}=\left(\sqrt{-g}\frac{ \delta \mathscr{L}}{ \delta
R_{\rho\mu\lambda\nu}}\right)^{\!\!(0)}\delta
R_{\rho\mu\lambda\nu}^{(2)}$.

We now evaluate $\delta R_{\rho\mu\lambda\nu}^{(2)}$. From~(\ref{expansion}) one finds that
\begin{equation}
\label{R2expansion}
\delta
R_{\rho\mu\lambda\nu}^{(2)}=\Upsilon_{\lambda\delta\rho}\Upsilon_{\nu\mu}^{\
\ \delta} -\Upsilon_{\delta\nu\rho}\Upsilon_{\mu\lambda}^{\ \
\delta}.
\end{equation}
Evaluating the product of $\Upsilon$'s gives
\begin{eqnarray}
\label{upsups}
\Upsilon_{\lambda\delta\rho}\Upsilon_{\nu\mu}^{\ \ \
\delta}&=&
\frac{1}{4}\left(\overline{\nabla}_{\lambda}h_{\delta\rho}+
\overline{\nabla}_{\delta}h_{\lambda\rho}-\overline{\nabla}_{\rho}h_{\lambda\delta}\right)
\left(\overline{\nabla}_{\nu}h_{\mu}^{\ \delta}+ \overline{\nabla}_{\mu}h_{\nu}^{\ \delta}-\overline{\nabla}^{\delta}h_{\nu\mu}\right) \nonumber \\
&=& \frac{1}{4} \left(
\overline{\nabla}_{\lambda}h_{\delta\rho}\overline{\nabla}_{\nu}h_{\mu}^{\
\delta} + \nabla_{\lambda}h_{\delta\rho}
\overline{\nabla}_{\mu}h_{\nu}^{\ \delta} -
\overline{\nabla}_{\delta}h_{\lambda\rho}\overline{\nabla}^{\delta}h_{\nu\mu}
- \overline{\nabla}_{\rho}h_{\lambda\delta}
\overline{\nabla}_{\nu}h_{\mu}^{\ \delta}
-\overline{\nabla}_{\rho}h_{\lambda\delta}
\overline{\nabla}_{\mu}h_{\nu}^{\ \delta} \right. \nonumber \\ &&
\left.-\overline{\nabla}_{\lambda}h_{\delta\rho}\overline{\nabla}^{\delta}h_{\nu\mu}+ \overline{\nabla}_{\delta}h_{\lambda\rho}\overline{\nabla}_{\nu}h_{\mu}^{\
\delta}+\overline{\nabla}_{\delta}h_{\lambda\rho}\overline{\nabla}_{\mu}h_{\nu}^{\delta}+ \overline{\nabla}_{\rho}h_{\lambda\delta}\overline{\nabla}^{\delta}h_{\nu\mu}
\right).
\end{eqnarray}
Substituting this expression into eq.(\ref{R2expansion}) and taking into account the symmetries of the Riemann tensor  $R_{\rho\mu\lambda\nu}$:  the symmetry under the double exchange $\rho\mu\leftrightarrow \lambda\nu$  and the antisymmetry under the exchanges $\rho\leftrightarrow\mu$, $\lambda\leftrightarrow \nu$ we find that
\begin{equation}
\label{R2expansionI}
\delta
R_{\rho\mu\lambda\nu}^{(2)}=\frac{1}{2}\left(\overline{\nabla}_{\delta}h_{\lambda\mu}\overline{\nabla}^{\delta}h_{\nu\rho} +2\overline{\nabla}^{\delta}h_{\lambda\rho}\overline{\nabla}_{\mu}h_{\nu\delta}\right).
\end{equation}
We can now explicitly exhibit the kinetic terms
\begin{eqnarray}
\label{final}
\delta I^{(2)}&&= \int d^D x\sqrt{-\overline{g}}\ \frac{1}{2}  \left(\frac{{\delta \mathscr{L}}}{ \delta R_{\rho\mu\lambda\nu}}\right)^{\!\!(0)} \left(\overline{\nabla}_{\delta}h_{\lambda\mu}\overline{\nabla}^{\delta}h_{\nu\rho} +2\overline{\nabla}^{\delta}h_{\lambda\rho}\overline{\nabla}_{\mu}h_{\nu\delta} \right).
\end{eqnarray}

It is possible to check in a straightforward manner that applying
the above procedure to the case of the Einstein-Hilbert action
reproduces exactly the result in eq.~(\ref{kappadef}).

\section{The Noether charge entropy is a quarter of the area in units of the effective gravitational coupling }
\label{noether}

By comparing eq.~(\ref{final}) to eq.~(\ref{waldentropygen})
\begin{align}
\delta I^{(2)}&= \int d^D x\sqrt{-\overline{g}}\ \frac{1}{2} \left(\frac{{\delta \mathscr{L}}}{ \delta R_{\rho\mu\lambda\nu}}\right)^{\!\!(0)} \left(\overline{\nabla}_{\delta}h_{\lambda\mu}\overline{\nabla}^{\delta}h_{\nu\rho} +2\overline{\nabla}^{\delta}h_{\lambda\rho}\overline{\nabla}_{\mu}h_{\nu\delta} \right) \cr
S_{W}&=-2 \pi \oint\limits_{\Sigma}  \left( \frac{\delta\mathscr{L}}{\delta R_{abcd}}\right)^{\!\!(0)} \hat\epsilon_{ab}\hat\epsilon_{cd}\bar{\epsilon},
\end{align}
we observe that the Noether charge formula involves the gravitational coupling of specific metric perturbation polarizations.  In the next section we show that these metric perturbations correspond to fluctuations of the area density on the bifurcation surface.

We may formally define
\begin{equation}
\label{kappadefgeneral}
\frac{1}{\left(\kappa_{eff}\right)^2}=-\frac{1}{4}\left( \frac{\delta\mathscr{L}}{\delta R_{abcd}}\right)^{\!\!(0)} \hat\epsilon_{ab}\hat\epsilon_{cd}.
\end{equation}
The factor $-1/4$ in eq.~(\ref{kappadefgeneral}) takes into account the symmetries of $R_{abcd}$ and the negative signature of the metric \cite{waldbook}.
Using definition (\ref{kappadefgeneral}) we find
\begin{equation}
\label{waldarea}
S_{W}= \frac{1}{4}\oint\limits_{\Sigma}\frac{32 \pi }{\left(\kappa_{eff}\right)^2}\bar{\epsilon}.
\end{equation}
In eq.~(\ref{waldarea}) the ``local unit of area"  $(32\pi)/\left(\kappa_{eff}\right)^2$ appears. It determines the weighting of the infinitesimal area bits in the integral. Identifying $G_{eff}=\frac{8 \pi}{\left(\kappa_{eff}\right)^2}$ we find
\begin{equation}
\label{waldareaI}
S_{W}= \frac{1}{4}\oint\limits_{\Sigma}\frac{dA}{G_{eff}}.
\end{equation}
If $G_{eff}$ is constant on the bifurcation surface then
\begin{equation}
S_{W}=\frac{A_H}{4 G_{eff}}.
\end{equation}

In the case of extremal BH's care should be taken when evaluating the effective coupling.
Wald's formula can be defined for extremal BH's since  it was shown in~\cite{jacobson} that the entropy can be
computed on any spatial section of the horizon and since the
entropy is a rescaled Noether charge in units in which the temperature is $1/2\pi$.
Therefore Wald's formula  applies to extremal black
holes if they are treated as limits of non-extremal ones. Similarly,  to define the effective coupling for extremal BH's and make the comparison with Wald's formula we have to treat extremal BH's as limits of non-extremal ones.

\section{The choice of  polarizations}
\label{polariz}

We have shown that the relevant kinetic terms originate from the second order expansion of the Riemann tensor,
\begin{equation}
\delta \mathscr{L}\sim
\left(\frac{{\delta \mathscr{L}}}{ \delta
R_{\rho\mu\lambda\nu}}\right)^{\!\!(0)}
\delta\,R_{\rho\mu\lambda\nu}^{(2)}.
\end{equation}
In Wald's formula a choice of specific polarizations of
$\delta\,R_{\rho\mu\lambda\nu}^{(2)}$ is made:
\begin{equation}
\label{choice}
\hat\epsilon^{ \rho \mu}\hat\epsilon^{\lambda \nu}\delta\,R_{\rho\mu\lambda\nu}^{(2)}.
\end{equation}
This choice is defined by the binormal vectors to the bifurcation surface.
Recall that on the bifurcation surface the Killing vector
$\tilde{\chi}^{b}$ vanishes and the binormal to the surface is given by
$\hat{\epsilon}_{ab}=\nabla_{a}\,\tilde{\chi_{b}}$.

We wish to identify the choice of polarizations in (\ref{choice}) with the fluctuations of the area density  $\hbox{\Large\it a}$ on the
bifurcation surface. The area of the bifurcation surface is
\begin{equation}
A_\Sigma=-\frac{1}{2}\oint\limits_{\Sigma}
\hat\epsilon^{ab}\hat\epsilon_{ab}\bar{\epsilon}.
\end{equation}
Since $\bar{\epsilon}$ is the induced volume form on the bifurcation
surface the area density can be defined as
\begin{equation}
\hbox{\Large\it a} =-\frac{1}{2} \hat{\epsilon}^{ab}\,\hat{\epsilon}_{ab}.
\end{equation}

Let us consider the following effective Lagrangian for the area density
\begin{equation}
\mathscr{L}_{\hbox{\Large\it a}}=\frac{1}{2} \hbox{\Large\it a}\
\nabla^2\hbox{\Large\it a}.
\label{effaaI}
\end{equation}
where $\sqrt{\hat{g}}$ is the determinant of the induced metric on
the bifurcation surface. Since
\begin{equation}
\nabla_{c}\hat{\epsilon}_{ab}=\nabla_{c}\nabla_{a}\tilde{\chi}_{b}=-R_{abcd}\tilde{\chi}^{d}
\end{equation}
we obtain
\begin{equation}
\nabla_{\sigma}\nabla_{c}\hat{\epsilon}_{ab}=-\nabla_{\sigma}\(R_{abcd}\tilde{\chi}^{d}\).
\label{lem1}
\end{equation}
On a bifurcation surface the Killing vector vanishes. It follows that
\begin{equation}
\nabla_{\sigma}\nabla_{c}\hat{\epsilon}_{ab}= -R_{abcd}\nabla_{\sigma}\tilde{\chi}^{d}=-R_{abc}^{\phantom{abc}d}\hat{\epsilon}_{\sigma d}.
\label{lem1}
\end{equation}
Substituting  eq.~(\ref{lem1}) into
\begin{equation}
\nabla^2(\hat{\epsilon}^{ab}\,\hat{\epsilon}_{ab})= 2\,\hat{\epsilon}^{ab}\,g^{\alpha\beta}\nabla_{\alpha}\nabla_{\beta}(\hat{\epsilon}_{ab})
\end{equation}
gives
\begin{equation}
\label{relation}
\nabla^2{\hbox{\Large\it a}}=
\,\hat{\epsilon}^{\alpha\beta}\,\hat{\epsilon}^{\gamma\delta}\,R_{\alpha\beta\gamma\delta}.
\end{equation}

Let us expand the Lagrangian (\ref{effaaI})  to second order.
In performing the expansion we use the fact that the normalization of the killing vector on the unperturbed bifurcation surface leads to $\hbox{\Large\it a}=1$ and we make a gauge choice such that
$\delta \hat{\epsilon}^{\alpha\beta}=0$ as in \cite{wald1}. The fluctuations of
the area density can be viewed as the difference in area density
between two slightly different surfaces. Let us denote the
difference in the metric between the perturbed and unperturbed
bifurcation surface by $\delta g_{\mu\nu}=h_{\mu\nu}$. Since we look
at two slightly different surfaces we have the freedom to
choose how the points in the two surfaces correspond. We will use
this freedom to make the correspondence such that the Killing vector
$\tilde{\chi^{a}}$ does not change from one surface to the other.
Thus $\delta\tilde{\chi}^{a}=0$ and\footnote{We do not have to
expand the covariant derivative since to first order in $h$ the
covariant derivative with respect to the background is equal to the
covariant derivative with respect to the perturbed metric.}
\begin{equation}
\delta \hat{\epsilon}^{\alpha\beta}
=\overline{\nabla}^{\alpha}\delta\,\tilde{\chi}^{\beta}=0.
\end{equation}

To summarize, we have shown that
\begin{equation}
\left(\hbox{\Large\it a}\
\nabla^2\hbox{\Large\it a}\right)^{(2)}=
\hat{\epsilon}^{\alpha\beta}\,\hat{\epsilon}^{\gamma\delta}\,\delta\,R_{\rho\mu\lambda\nu}^{(2)}.
\label{almostfinal}
\end{equation}
In other words, we have  shown that the specific polarization of gravitons that appears
in the expansion of (\ref{effaaI}) is the same one that appears in Wald's
entropy formula.

\section{Examples }
\label{examples}

In this section we present three examples. The purpose of presenting the first two examples is to check explicitly the proposed relationship between the gravitational coupling and the functional derivative of $\mathscr{L}$ with respect to the Riemann tensor. We do this by expanding $\mathscr{L}$ to second order. The third example shows that the relationship between $G_N$ and $G_{eff}$ can be non-analytic in certain cases and resolves a long-standing puzzle \cite{dabholkar} as to why in $N=2$ SUGRA BH's (and in small heterotic BH's)  $S=A/2 G_N$ rather than $A/ 4 G_N$.

\subsection { $R+\lambda R^n$}

As a first example let us consider the following Lagrangian
\[\mathscr{L}=\frac{1}{16\pi G_N}\left(R+\lambda R^n\right).\] The
calculation of Wald's Noether charge entropy gives (substitution in
eq.~(\ref{waldexplicit})):
\begin{eqnarray}
\label{rrn}
S_{W}&=& -\frac{1}{4 G_N} \oint\limits_{r=r_H, t=const.}   \left(\overline{g}^{tt}\overline{g}^{rr}- \overline{g}^{rt}\overline{g}^{rt}\right) \left(1+n \lambda  R^{n-1}\right)\left[ {q}(r)\right]^{\frac{d-1}{2}} d\Omega_{d-1}^2  \nonumber \\
&=& \frac{{A}_H}{4\left(1+n\lambda
\left[R(r_H)\right]^{n-1}\right)^{-1} G_N}.
\label{Waldexample1}
\end{eqnarray}

Again, in this case $G_N$ does not determine by itself the strength
of the gravitational coupling or of the gravitational force. The
similarity to the case of dilaton-gravity can be made more explicit
by performing a field redefinition into the Einstein frame
\cite{jacobson}. The gravitational coupling of the specific metric perturbation
polarization $(t,r)$ can be obtained using eq.~(\ref{kappadefgeneral}):
\begin{eqnarray}
\label{kappaexample1}
\frac{1}{\left(\kappa_{eff}\right)^2}&=&-\frac{1}{2}\frac{1}{16\pi G_N}\left(\overline{g}^{tt}\overline{g}^{rr}- \overline{g}^{rt}\overline{g}^{rt}\right) \left(1+n \lambda  R^{n-1}\right) \nonumber \\ &=&\frac{1}{32\pi G_N}\(1+\lambda n {R}^{n-1}\).
\end{eqnarray}
In this case the entropy in eq.~(\ref{rrn}) becomes
$$
S_W=\frac{A_{H}}{4G_{eff}}.
$$

The computation of the kinetic term for this example is as
follows. Using the expansion of $R$ to second order in
$h_{\mu\nu}$ from section \ref{kinetic}:
\begin{equation}
R=\overline{R}+\frac{1}{4}\left(\,\overline{\nabla}_{\alpha}\,h_{\mu\nu}\, \overline{\nabla}^{\alpha}\,h^{\mu\nu}- 2\overline{\nabla}_{\alpha}\,h_{\mu\beta}\,\overline{\nabla}^{\beta}\,h^{\mu\alpha} -\overline{\nabla}_{\alpha}h\, \overline{\nabla}^{\alpha}h+ 2 \overline{\nabla}_{\alpha} h\overline{\nabla}_{\beta}\,h^{\alpha\beta}  \right)+\cdots,
\label{expansion of R}
\end{equation}
we obtain
\begin{eqnarray}
&&\frac{1}{16\pi G_N}\left(R+\lambda
R^n\right)=\frac{1}{16\pi G_N}\(\overline{R}+\lambda
\overline{R}^n\) \cr &+&\frac{1}{64\pi
G_N}\,\(1+\lambda\,n\,\overline{R}^{n-1}\)\, \left(\,\overline{\nabla}_{\alpha}\,h_{\mu\nu}\, \overline{\nabla}^{\alpha}\,h^{\mu\nu}- 2\overline{\nabla}_{\alpha}\,h_{\mu\beta}\,\overline{\nabla}^{\beta}\,h^{\mu\alpha} -\overline{\nabla}_{\alpha}h\, \overline{\nabla}^{\alpha}h+ 2 \overline{\nabla}_{\alpha} h\overline{\nabla}_{\beta}\,h^{\alpha\beta}  \right)
\nonumber
\end{eqnarray}
so that we identify the prefactor of the kinetic term
\begin{equation}
\kappa_{eff}^2=32\,\pi\,G_N\,\(1+\lambda\,n\,\overline{R}^{n-1}\)^{-1},
\end{equation}
and thus
\begin{equation}
G_{eff}=G_N\,\(1+\lambda\,n\,\overline{R}^{n-1}\)^{-1}.
\end{equation}
This is the same $G_{eff}$ that we obtained in eq.~(\ref{kappaexample1}).

\subsection {$R+\lambda\,R_{\rho\mu\lambda\nu}\,R^{\rho\mu\lambda\nu}$}

Next let us consider a more complicated example:
\[\mathscr{L}=\frac{1}{16\pi G_N}\left(R+\lambda\,R_{\rho\mu\lambda\nu}\,R^{\rho\mu\lambda\nu}\right).\]
Since Wald's formula is linear in the Lagrangian we can
substitute $\frac{\lambda}{16\pi G_N}\,R_{\rho\mu\lambda\nu}\,R^{\rho\mu\lambda\nu}$ in eq.~(\ref{waldexplicit}) and obtain the correction term to the Bekenstein-Hawking entropy,
\begin{equation}
 -8 \pi \oint\limits_{r=r_H, t=const.}
\frac{\lambda}{8\,\pi\,G_N}R^{rtrt}\left[q(r)
\right]^{\frac{d-1}{2}} d\Omega_{d-1}^2= -\frac{\lambda
\,R^{rtrt}}{G_N}\,A_H.
\end{equation}
Then
\begin{equation}
S_{W}=\frac{A_{H}}{4\,G_N}\(1-4\,\lambda\,R^{rtrt}\).\label{Waldexample2}
\end{equation}
The gravitational coupling of the specific metric perturbation polarization
$(t,r)$ can be obtained using eq.~(\ref{kappadefgeneral}):
\begin{eqnarray}
\label{kappaexample2}
\frac{1}{\left(\kappa_{eff}\right)^2}&=&-\frac{1}{2}\frac{1}{16\pi G_N}\left(\overline{g}^{tt}\overline{g}^{rr}- \overline{g}^{rt}\overline{g}^{rt}+4\lambda R^{rtrt}\right) \nonumber \\
&=&\frac{1}{32\pi G_N}\(1-4\lambda R^{rtrt}\).
\end{eqnarray}
In this case the entropy becomes
\be \label{entropyexample2}
S=\frac{A_{H}}{4 G_{eff}}.
\ee

On the other hand the expansion of $\frac{\lambda}{16\pi
G_N}\,R_{\rho\mu\lambda\nu}\,R^{\rho\mu\lambda\nu}$ to second
order\footnote{The contribution of the expansion of
$R_{\rho\mu\lambda\nu}$ to first order in $h_{\mu\nu}$ vanishes
according to the general discussion in section \ref{kinetic}.} in
$h_{\mu\nu}$ according to eq.~(\ref{R2expansionI}), is the following
\begin{equation}
\frac{\lambda}{8\pi
G_N}\,\left(\Upsilon_{\lambda\delta\rho}\Upsilon_{\nu\mu}^{\
\ \delta} -\Upsilon_{\delta\nu\rho}\Upsilon_{\mu\lambda}^{\ \
\delta}\right)\,R^{\rho\mu\lambda\nu}=\frac{\lambda}{16\pi G_N}\,\left(\overline{\nabla}_{\delta}h_{\lambda\mu}\overline{\nabla}^{\delta}h_{\nu\rho}
+2\overline{\nabla}^{\delta}h_{\lambda\rho}\overline{\nabla}_{\mu}h_{\nu\delta} \right)\,R^{\rho\mu\lambda\nu} \ \ \ \ \
\end{equation}
When we add this contribution to the EH action
(\ref{expansion of R}) we get the full kinetic term
\begin{eqnarray}
\frac{1}{64\,\pi\,G_{N}}\,\( g^{\rho\lambda} g^{\mu\nu}-g^{\rho\nu}g^{\mu\lambda} -4\,\lambda\,R^{\rho\mu\lambda\nu}\) \left(\overline{\nabla}_{\delta}h_{\lambda\mu}\overline{\nabla}^{\delta}h_{\nu\rho}
+2\overline{\nabla}^{\delta}h_{\lambda\rho}\overline{\nabla}_{\mu}h_{\nu\delta}
\right).
\end{eqnarray}
In this example we have to take the $(t,r)$ sector and then we obtain
\begin{equation}
\kappa_{eff}^2=32\,\pi\,G_N\,\(1-4\,\lambda\,R^{trtr}\)^{-1}, \ee
and thus
\begin{equation}
G_{eff}=G_N\,\(1-4\,\lambda\,R^{trtr}\)^{-1}.
\end{equation}
This is the same $G_{eff}$ that we obtained in eq.~(\ref{kappaexample2}).

\subsection {Small black holes in heterotic string theory}

Sen has defined  the BH entropy
function \cite{sen-function} and used it  \cite{sen2005,sen2007} to find the near horizon geometry of extremal BH solutions in the four
dimensional low energy effective action of heterotic string theory
on $\cal {M}$ $ \times S^{1} \times \widetilde{S^{1}}$. The manifold $\cal
M$ is some four manifold suitable for heterotic string compactification. The entropy function can be used to compute the entropy of such BH's.

For completeness we recall the field content and action of Sen's construction.
The four dimensional fields relevant for the construction of this
are related to the ten dimensional string metric $G_{MN}^{(10)}$,
anti-symmetric tensor field $B_{MN}^{(10)}$ and the dilaton
$\Phi^{(10)}$ via the relations:
\begin{equation}
\begin{array}{llll}
\Phi &= \Phi^{(10)} - {1 \over 4} \, \ln (G^{(10)}_{99})
 - {1 \over 4} \, \ln (G^{(10)}_{88}) - {1 \over 2}\ln V_{\cal M}\, ,  & &
 \\
  G_{\mu\nu} & = G^{(10)}_{\mu\nu} - (G^{(10)}_{99})^{-1} \,
G^{(10)}_{9\mu} \, G^{(10)}_{9\nu} - (G^{(10)}_{88})^{-1} \,
G^{(10)}_{8\mu}
\, G^{(10)}_{8\nu}\, ,&&  \\
 S &=e^{-2\Phi}\, ,
  &\hspace{-3.1in} R  = \sqrt{G^{(10)}_{99}}\, ,
 &\hspace{-2.1in}\widetilde{ R}  = \sqrt{G^{(10)}_{88}}\, ,  \\
 A^{(1)}_\mu &= {{1 \over 2}} (G^{(10)}_{99})^{-1} \,
G^{(10)}_{9\mu}\, , &\hspace{-2.2in} A^{(2)}_\mu = {{1 \over 2}}
(G^{(10)}_{88})^{-1} \, G^{(10)}_{8\mu}\, ,&
 \\
A^{(3)}_\mu  &= {1 \over 2} B^{(10)}_{9\mu}\, ,
& \hspace{-2.2in} A^{(4)}_\mu ={1 \over 2} B^{(10)}_{8\mu},&
\end{array}
\end{equation}
where $V_{\cal M}$ denotes the volume of $\cal M$ measured in the string metric.
The effective action of these fields is given by
\begin{eqnarray}
I &=& \frac{1}{ 16\pi G_N} \int d^4 x \, \sqrt{-\det G} \, S \, \bigg[
R_G
+ S^{-2}\, G^{\mu\nu} \, \partial_\mu S \partial_\nu S \nonumber \\
&-&  R^{-2} \, G^{\mu\nu} \, \partial_\mu R \partial_\nu R -
\widetilde{R}^{-2}
\, G^{\mu\nu} \, \partial_\mu \widetilde{R} \partial_\nu \widetilde{R} \nonumber \\
&-& R^2 \, G^{\mu\nu} \, G^{\mu'\nu'} \, F^{(1)}_{\mu\mu'}
F^{(1)}_{\nu\nu'} - \widetilde{R}^2 \, G^{\mu\nu} \, G^{\mu'\nu'} \,
F^{(2)}_{\mu\mu'} F^{(2)}_{\nu\nu'}
\nonumber \\
&-& R^{-2} \, G^{\mu\nu} \, G^{\mu'\nu'} \, F^{(3)}_{\mu\mu'}
F^{(3)}_{\nu\nu'}
 - \widetilde{R}^{-2} \,
G^{\mu\nu} \, G^{\mu'\nu'} \, F^{(4)}_{\mu\mu'}
F^{(4)}_{\nu\nu'}\bigg] \nonumber \\
&+&\hbox{ higher derivative terms + string loop corrections}.
\end{eqnarray}

In \cite{sen2005} Sen considered extremal BH's with two electric and two magnetic charges
assuming the near horizon solution is of the form $AdS_2\times S^2$:
\begin{equation}
    ds^2=\nu_1\left[-r^2dt^2+\frac{dr^2}{r^2}\right]+\nu_2d\Omega^2.
\label{ansatz}
\end{equation}
Here $\nu_1$, $\nu_2$ and all the additional fields are
constant on the horizon. By extremizing of the entropy
function one obtains the solution (justifying the ansatz above):
\begin{eqnarray}
\nu_1=\nu_2=4\,N\,W,\\
e^{-2\,\phi(r_H)}=\sqrt{\frac{nw}{NW}},
\end{eqnarray}
where $n,w$ are electric charges that correspond to momentum and
winding modes of the fundamental string, $N,W$ are the corresponding
magnetic charges and $\phi(r_H)$ is the value of the dilaton field
on the horizon. Since the solutions are extremal they may be
expressed as a function of the charges without an explicit
dependence on the mass.

The area of the horizon is
\be
 A=4\,\pi\,\nu_2=16\,\pi\,N\,W,
\ee
and the entropy is, according to eq.~(\ref{entropy-dilaton}),
\be
S_W=\frac{A}{4\,G_N}\,e^{-2\,\phi(r_H)}=\frac{4\,\pi}{G_N}\sqrt{n\,w\,N\,W}.
\ee

Such BH's are singular in the limit when the magnetic charges, either $N$ or
$W$, go to zero. In this case the horizon area vanishes and
consequently also the entropy. In this limit $\alpha'$
corrections become important since the curvature is large in
comparison with the radius of the BH.
To model the effects of the corrections Sen considered the addition of a Gauss-Bonnet term to the original low energy effective Lagrangian:
\[\frac{\lambda}{16\pi\,G_{N}}\,e^{-2\phi}\left[R_{\mu\nu\rho\sigma}R^{\mu\nu\rho\sigma}-4R_{\mu\nu}R^{\mu\nu}+R^2\right]\]
where $\lambda$ is equal to $\alpha'$ up to a numerical constant.
Using the same ansatz (\ref{ansatz}) for the near horizon solution
Sen obtained the following solution:
\bea
\label{nusen}
\nu_1=\nu_2=4\,N\,W+\frac{8\,\lambda}{G_N}\\
\label{phisen}
e^{-2\,\phi(r_H)}=\sqrt{\frac{nw}{NW+4\frac{\lambda}{G_N}}}.
\eea
 The area of the horizon in this case is
\be
\label{area sen}
 A=4\,\pi\,\nu_2=16\,\pi\,N\,W+32\,\pi\,\frac{\lambda}{G_N}.
 \ee

The gravitational coupling of the specific metric perturbation polarization $(t,r)$, can be obtained using eq.~(\ref{kappadefgeneral})
\begin{eqnarray}
\label{kappaexample3A} \frac{1}{\left(\kappa_{eff}\right)^2}=
&-&\frac{1}{64\pi G_N}e^{-2\phi}\left[2\overline{g}^{tt}\overline{g}^{rr}+2\lambda \left(4 R^{rtrt}-4 \overline{g}^{tt}R^{rr}-4 \overline{g}^{rr}R^{tt}+2\overline{g}^{tt}\overline{g}^{rr}R\right)\right]= \nonumber \\
&=&\frac{1}{32\pi G_N}e^{-2\phi}\left[1+2\lambda
\left(2R^{rt}_{\;\;\;rt}- 2\overline{g}_{rr}R^{rr}-
2\overline{g}_{tt}R^{tt}+R\right)\right].
\end{eqnarray}
Using the metric (\ref{ansatz}) and the solution (\ref{nusen}),
(\ref{phisen}) one gets that the gravitational coupling of the
specific metric perturbation polarization $(t,r)$ is
\be
G_{eff}=G_N\,\frac{\,N\,W+\frac{2\,\lambda}{G_N}}{\sqrt{n\,w\,\(NW+4\,\frac{\lambda}{G_{N}}\)}},
\ee
and the entropy becomes
\be \label{entropy sen}
\frac{A}{4G_{eff}}=\frac{4\,\pi}{G_{N}}\sqrt{n\,w\,\(NW+4\,\frac{\lambda}{G_{N}}\)}.
\ee
In the limit of a small BH we obtain that \be
G_{eff}=\sqrt{\frac{\lambda\,G_N}{n\,w}}=\frac{G_N}{2}e^{2\phi}. \ee
This example is different than the previous ones since the
dependence of the effective coupling  is not analytic in
$G_N$. This may be expected due to the singular behavior of the
horizon in the original solution which is resolved by the added
Gauss-Bonnet term.

Transforming to the Einstein frame we get $A_E=A\,e^{-2\,\phi}$
 and therefore for small black holes we obtain that
\be
G_{eff}=\frac{G_N}{2}.
\ee
The same result is, of course, obtained by direct application of eq.~(\ref{kappadefgeneral})
and thus the entropy becomes
\be
  \label{entropy sen1}
\frac{A_E}{4G_{eff}}=\frac{A_E}{2G_N}.
\ee
The factor of 2 difference between Wald's entropy and the Bekenstein-Hawking entropy has been somewhat of a puzzle since its discovery in the context of $N=2$
SUGRA~\cite{dabholkar}. However, it is simply explained by the difference in the effective gravitational coupling.

\section{Conclusions and outlook}
\label{discussion}

We have found that the Noether charge entropy is equal to an integral over the horizon of the ``entropy density" $dS_W=\frac{dA}{4\,G_{eff}}$. The only difference between the Wald entropy and the Bekenstein-Hawking entropy is that the ``unit of area" rather than being Newton's constant $G_N$ is $G_{eff}$.  We believe that this simple appealing expression may be valid for a more general class of black holes, not only those for which the Noether charge entropy can be defined.

The $G_{eff}$ in Wald's entropy is associated with a specific metric perturbation polarization. We have been able to identify the polarizations with area fluctuations on the bifurcating surface. It would be interesting to relate the choice of polarization to the fact that the entropy satisfies the first law and to understand the choice from a dynamical point of view. Perhaps this polarization is related to the response of the black hole to a change in its energy.

We have been able to verify our proposal only for static backgrounds, which in our formulation are spherically symmetric. For spherically symmetric solutions the effective coupling is trivially constant on the horizon. Stationary (non-static) solutions may involve a varying effective coupling, so it would be interesting to find such solutions and to put our proposal to a non-trivial test. For this we would need an example of a stationary black hole solution (including higher derivative corrections) in string theory whose entropy can be calculated via microstates counting. We were not able to find any such solutions in the literature.

The local and observer dependent expression for the entropy is consistent with the entanglement interpretation of BH entropy and hence resolves the apparent tension between the Wald's entropy of BH's in higher derivatives theories and the entanglement entropy \cite{waldliving}. The entanglement ``entropy density" has the form $d\,S_{Entanglement}=dA/4\delta^{D-2}$ with $\delta^{D-2}$ being some ``unit of area" defined by a UV scale in the theory. Our results suggest that $\delta^{D-2}$ should be proportional to $G_{eff}$.

Our result explains in a simple way the results of \cite{giveon1,giveon2} where it was found that entropy of certain BH's is proportional to the area as a function of the charges rather than being a more general function of the charges. We now understand that in the examples discussed in \cite{giveon1,giveon2} the effective gravitational coupling $G_{eff}$ is determined only by the dilaton on the horizon which is independent of the charges.

Our results should be extendible to cosmological spacetimes. It has long been suspected that entropy bounds may provide important clues to the nature of cosmological singularities and their possible resolution. The form of the entropy bounds in theories with higher derivatives has been under debate (For a review, see for example \cite{Brustein:2007hd}). Our result suggests a specific form for cosmological entropy bounds.

\section{Acknowledgments}
We thank Atish Dabholkar for discussions, Amit Giveon for useful suggestions and
comments and Ted Jacobson for comments on the manuscript and useful suggestion. We thank Bob Wald for useful questions and comments on the manuscript.

The research of RB and MH was supported by The Israel Science Foundation grant no 470/06. MH research was supported by The Open University of Israel's Research Fund .
DG thanks the National Sciences and Engineering Research Council of Canada for the financial support. Part of this work was done while DG was in the Hebrew University of Jerusalem, where he was supported by The Israel Science Foundation grant no 607/05,  DIP grant H.52, EU grant MRTN-CT-2004-512194 and the Einstein Center at the Hebrew University.

\end{document}